\documentclass{emulateapj}
\usepackage{apjfonts}
\usepackage{graphicx}

\slugcomment{to appear in the Astrophysical Journal Letters}

\shorttitle{The neutron star in V395~Car/2S~0921--630}
\shortauthors{Steeghs \& Jonker}

\begin{document}


\title{On the mass of the neutron star in V395~Car/2S~0921--630}


\author{D. Steeghs\altaffilmark{1,2}}
\email{D.T.H.Steeghs@warwick.ac.uk}
\altaffiltext{1}{Department of Physics, University of Warwick, Coventry, CV4 9BU, UK}
\altaffiltext{2}{Harvard-Smithsonian Center for Astrophysics, 60 Garden Street,
Cambridge, MA~02138, Massachusetts, U.S.A.}
\author{P. G. Jonker\altaffilmark{2,3,4}}
\altaffiltext{3}{SRON, Netherlands Institute for Space Research, Sorbonnelaan 2, 3584 CA, Utrecht, NL}
\altaffiltext{4}{Astronomical Institute, Utrecht University, P.O.Box 80000, 3508 TA, Utrecht, NL}




\begin{abstract}

We report high-resolution optical spectroscopy of the low-mass X-ray binary V395
Car/2S~0921--630 obtained with the MIKE echelle spectrograph on the Magellan-Clay telescope.
Our spectra are obtained near inferior conjunction of the mass donor star and we exploit the absorption lines originating from the back-side of the K-type object to accurately derive its rotational velocity. Using K0-K1 III templates, we find $v\sin{i}=32.9\pm 0.8$ km s$^{-1}$. We show that the choice of template star and the assumed limb darkening coefficient has little impact on the derived rotational velocity.
This value is a significant revision downwards compared to previously published values. We derive new system parameter constraints in the light of our much lower rotational velocity. We find $M_1=1.44 \pm 0.10  M_{\odot}$, $M_2=0.35 \pm 0.03  M_{\odot}$, and $q=0.24 \pm 0.02$ where the errors have been estimated through a Monte-Carlo simulation.
A possible remaining systematic effect is the fact that we may be over-estimating the orbital velocity of the mass donor due to irradiation effects. However, any correction for this effect will only reduce the compact object mass further, down to a minimum mass of $M_1=1.05 \pm 0.08  M_{\odot}$.
There is thus strong evidence that the compact object in this binary is a neutron star of rather typical mass and that the previously reported mass values of 2-4$M_{\odot}$ were too high due to an over-estimate of the rotational broadening.

\end{abstract}





\section{Introduction}

\noindent Low--mass X--ray binaries (LMXBs) are binary systems in which a
neutron star or a black hole accretes matter from a low--mass companion star.
These systems provide a laboratory to test the behavior of matter under
physical conditions that are unattainable on Earth. One of the ultimate
goals of the study of neutron stars is to determine the equation of state (EoS) that describes the relation  between
pressure and density of matter under the extreme conditions encountered in
neutron stars \citep{2001ApJ...550..426L,2004Sci...304..536L}.
\noindent As a result of the current paucity of observational constraints there
are many theories of the EoS of matter at neutron--star densities. Through the 
measurement of the masses and radii of neutron stars these theories can be tested.
For a specific EoS, one can derive a firm upper limit on the mass of the neutron star, above
which the object is not stable and would collapse into a black hole.  Neutron
stars with masses well above $1.4\,M_\odot$ cannot exist for so--called soft
EoSs. Therefore, measuring a high mass for even one neutron star would imply
the firm rejection of many proposed EoSs (see discussion by
\citealt{1995xrb..book...58V}).

V395~Car/2S~0921--630 is a promising accreting binary system for accurate system parameter work. It shows regular eclipses in the X-rays \citep{1987MNRAS.226..423M} as well as dips in the optical lightcurve \citep{1981A&A....94L...3C}. These reveal a 9.02d orbital period and imply a high orbital inclination. The unknown orbital inclination is often a significant handicap when trying to determine binary parameters, but in V395 Car the $\sin{i}$ factor is well constrained. In the optical, the K-type donor star absorption line spectrum is visible \citep{1983MNRAS.205..403B} which allows us to determine its rotational velocity, $v\sin{i}$  \citep{1999A&A...344..101S}, as well as the radial velocity amplitude $K_2$ (\citealt{2004ApJ...616L.123S,2005MNRAS.356..621J}).
These previous studies suggested the presence of a rather massive compact object with a mass between $\sim$2-4$M_{\odot}$. In the case of V395~Car the main source of
error in the mass of the compact object is the large error on the rotational
velocity (Jonker et al.~2005) which was derived from relatively low resolution data \citep{1999A&A...344..101S}. 

Here, we present Magellan Inamori Kyocera Echelle (MIKE) data of the V$\sim$16 magnitude
counterpart of V395 Car/2S~0921--630 to determine the rotational velocity of
the companion star accurately. 
We show in this {\it Letter} that we find a significantly lower value for the rotational broadening than previously determined, and explore the effects on the inferred neutron star mass.

\section{Observations \& reduction}

We observed V395~Car/2S~0921--630 with the MIKE echelle spectrograph mounted on
the Magellan Clay telescope at Las Campanas Observatory. Four exposures of 30 mins each were obtained on 
Jan.~25, 2005 (MJD 53395) between 7:05-9:15 UTC. We employed the spectrograph in its dual-beam mode using a 1" wide slit and a dichroic. The 2048x4096 pixel CCD detectors were binned on-chip in 2x2 mode. This delivers a wavelength coverage of 3430-5140\AA~on the blue MIT detector with a spectral dispersion of 0.04\AA/pixel, while the red SITe detector covered 5220-9400\AA~at 0.10\AA/pixel.  The spectrophotometric standard HR 4468 was observed immediately after the target exposures. Observing conditions were good with clear skies and 0.5" seeing which resulted in a seeing limited resolution of $\sim$0.08\AA~in the blue and 0.17\AA~in the red. Exposures of 5 bright template stars of spectral type K were obtained with the same setup a few nights earlier on Jan.~21 2005.

The data were reduced using the \textsc {MIKE redux}\footnote{http://web.mit.edu/$\sim$burles/www/MIKE/} software package that is tailored for the instrument. It performs the standard CCD calibration steps, constructs a wavelength solution using ThAr exposures obtained at the position of the target and optimally extracts the object frames. The orders were merged using our spectrophotometric flux standard and then exported into our analysis software (\textsc {molly}).
We rebinned all our object and template star spectra onto the same wavelength scale using a uniform velocity scale and their continua were normalised using spline fits while masking strong spectral features. All spectra were shifted into a heliocentric velocity frame by removing the Earth's velocity.

\section{Analysis}

\begin{table}
\begin{center}
\caption{Rotational broadening as a function of template\label{tab-temp}}
\begin{tabular}{llcc}
\tableline\tableline
Star & Spectral & ~~~~~~~~~~~~$v\sin{i}$ (km s$^{-1}$)$^{a}$ & \\
 ID    &   type       & 0.5  & 0.75\\     
\tableline
HD~100623 & K0V   & 30.5$\pm$0.8 & 31.5$\pm$0.9\\
HD~124106 & K1V   & 30.5$\pm$0.8 & 31.4$\pm$0.8\\
HD~99322  & K0III & 32.3$\pm$0.7 & 33.6$\pm$0.8\\
HD~121416 & K1III & 31.3$\pm$0.7 & 32.5$\pm$0.8\\
HD~83240  & K1III & 31.4$\pm$0.8 & 32.7$\pm$0.9\\
\tableline
MS mean & & 30.5$\pm$0.8 & 31.5$\pm$0.9\\
Giant mean & & 31.7$\pm$0.7 & 32.9$\pm$0.8\\
\tableline
\end{tabular}
\tablenotetext{a}{The two columns reflect our two choices for the linear limb darkening coefficient}
\end{center}
\end{table}

The principle aim of our observations was to accurately determine the rotational velocity of the mass donor star.  We intentionally obtained the observations close to inferior conjunction of the mass donor star
when its contribution to the overall light is expected to be the largest. Another advantage is that at this phase we observe the back side of the donor star. This side is not subjected to X-ray heating due to the luminous accretion flow around the compact object and the non-spherical effects due to its Roche-lobe shape are minimal. According to the ephemeris of Jonker et al. (2005), our observations span 0.89-0.90 in orbital phase.

\begin{figure}
\centerline{\includegraphics[angle=-90,width=8cm]{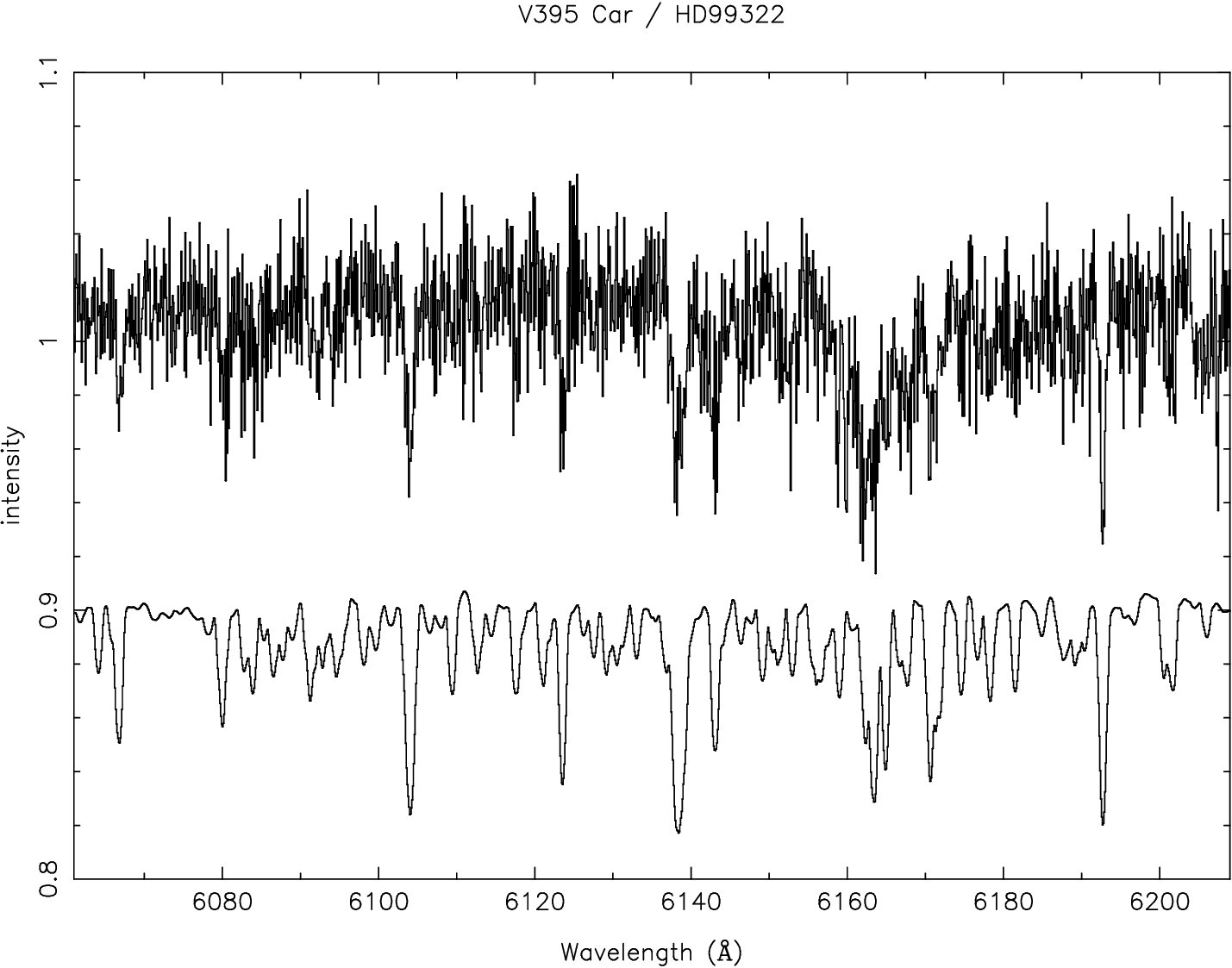}}
\centerline{\includegraphics[angle=-90,width=8cm]{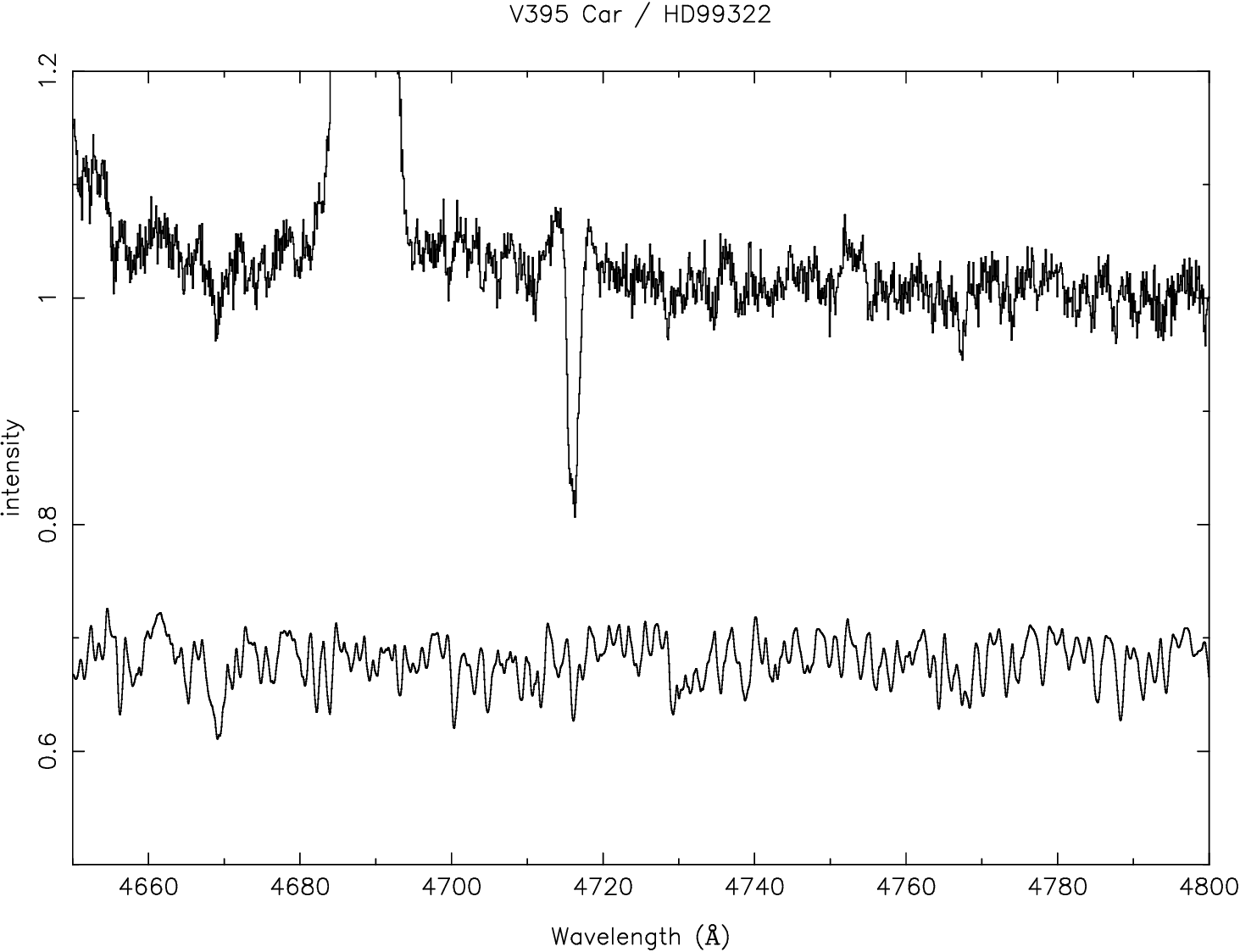}}
\caption{Two small segments of data showing the mean spectrum of V395 Car together with a broadened K0III template. A large number of resolved absorption lines that are well matched by early K-stars are clearly visible in both wavelength regions. The bottom panel is from the blue arm data and also contains the strong HeII 4686 emission line as well as HeI 4713 absorption originating in the accretion flow .\label{figspec}}
\end{figure}

Any intrinsic velocity shifts between our target spectra and the template stars were first removed by cross-correlating the normalised spectra while masking the emission lines in our target spectra as well as regions containing telluric absorption features. The template stars were then rotationally broadened by a varying amount using a spherical model and including a linear limb darkening coefficient. We then subtracted these broadened templates from our individual target spectra while optimising for a variable scaling factor $f$ reflecting the fact that not all the light in the system originates from the mass donor. This optimal subtraction is performed by minimizing the root mean square (rms) residuals of the data -
($f\times$constant + (1-$f$)$\times$ template spectrum). Again the regions containing telluric absorption and H and He emission lines in the object spectrum were masked. The goodness of fit of this subtraction is then evaluated as a function of the applied rotational broadening using the $\chi^2$-statistic. We plot two small segments of our observed average spectrum of V395 Car together with a broadened template spectrum in Figure~\ref{figspec}. 

\begin{figure}

\centerline{\includegraphics[angle=-90,width=7.8cm]{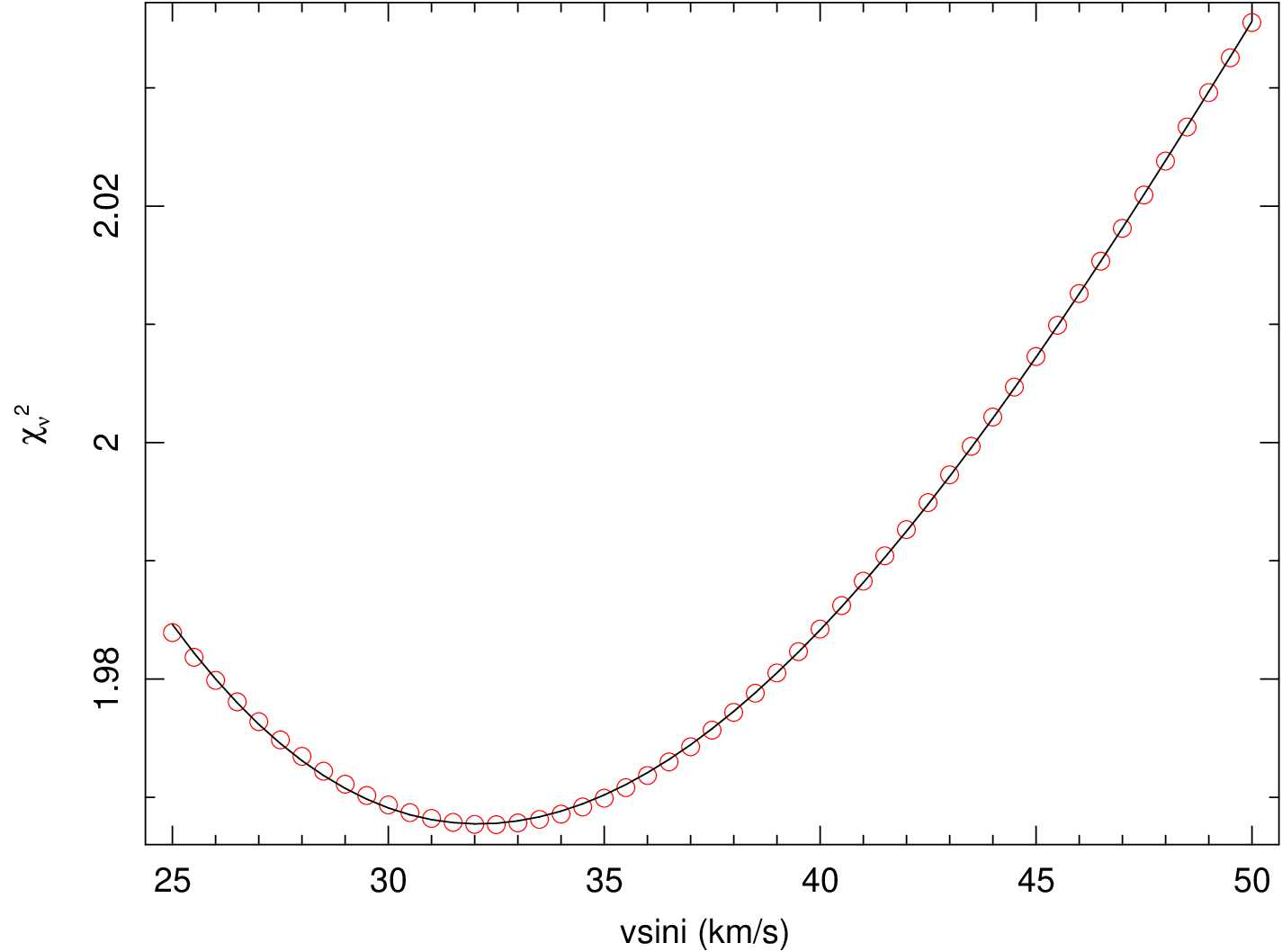}}
\centerline{\includegraphics[angle=-90,width=8cm]{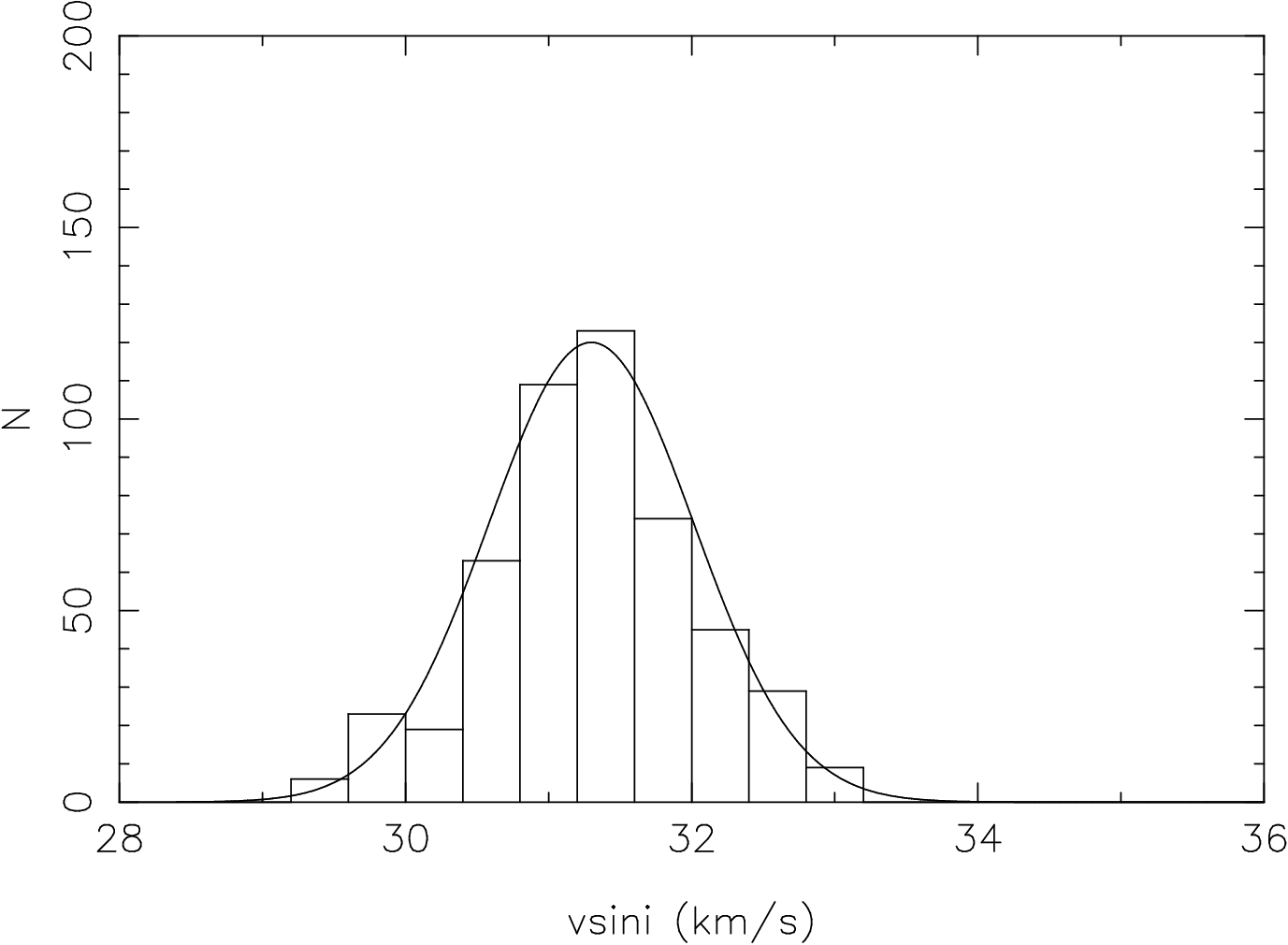}}

\caption{{\it a)} An example showing the dependence of the achieved $\chi_{\nu}^2$ statistic for an optimally subtracted template star as a function of the applied rotational broadening $v\sin{i}$. Symbols show the actual values returned from our fits, while the solid line is a cubic fit that is used to determine the best $v\sin{i}$ for each run.
{\it b)} Histogram of the derived rotational broadening values calculated from 500 trial runs of subtracting the K1III star HD 121416. The values are well characterised by a Gaussian distribution as shown by the model Gaussian and provides the final $v\sin{i}$ and its error for a given template, 31.3$\pm$0.7 km s$^{-1}$ in this case.\label{figoptsini}}
\end{figure}

We paired up the individual spectra with our 5 templates and found that in all cases, the optimal rotational broadening was found to be close to 32 km s$^{-1}$, a value that is significantly lower than the $64\pm9$ km s$^{-1}$ reported by Shahbaz et al. (1999). We found no significant differences between the values derived from the individual exposures and thus averaged our four shifted spectra together using variance weights to boost the S/N. We also found that the red spectra are preferred since they contain wide spectral windows free from emission lines where strong K-star absorption features are expected. For our final analysis, we therefore repeated our rotational broadening analysis using the mean red spectrum of V395 Car and broadening all templates between 25-50 km s$^{-1}$ in steps of 0.5 km s$^{-1}$. The optimal $v\sin{i}$ value for each template and for a given choice of the limb darkening coefficient was derived by fitting a cubic function to the $\chi^2$ versus $v\sin{i}$ values (Figure~\ref{figoptsini}a). 
Our best fit solution usually delivered a $\chi_{\nu}^2\sim2$ reflecting the rather simplistic assumption that the spectrum can be modeled by a constant and featureless accretion disk spectrum and the mass donor star as represented by our spectral templates. In order to determine more reliable error estimates on our $v\sin{i}$ values, we used a Monte-Carlo approach. We simulated 500 copies of our input target spectrum using a bootstrap technique where the input spectrum is resampled by randomly selecting data points from it. Each bootstrap copy always has the same number of data points as the input spectrum by allowing points to be selected multiple times or not at all. For each bootstrap copy, the optimal $v\sin{i}$ is evaluated as described above. We found that the $v\sin{i}$ values were well described by a Gaussian distribution (Figure~\ref{figoptsini}b), and thus calculated the mean and rms of our 500 $v\sin{i}$ values to deliver a $v\sin{i}$ value and its 1$\sigma$ error for each template star. We provide the results for our 5 template stars in Table~\ref{tab-temp} for two values of the limb darkening coefficient. For K-giants, one expects a typical limb darkening coefficient of 0.75 in the continuum \citep{1995A&AS..114..247C}, but we also list the equivalent values for an assumed limb darkening coefficient of 0.50 in order to determine its effect on the derived $v\sin{i}$ values.

\section{Implications}

Thanks to the high spectral resolution and good S/N of our Magellan data, we were able to extract accurate $v\sin{i}$ values with $<1$ km s$^{-1}$ errors. The value was also stable against choice of template, although the required $v\sin{i}$ for the giant templates is on average slightly higher than that for the main-sequence templates (Table~\ref{tab-temp}). This likely reflects the fact that the absorption lines found in the unbroadened main sequence templates have higher pressure atmospheres and thus are intrinsically broader and require a slightly smaller amount of additional rotational broadening. This difference is $<2\sigma$ as is the difference between individual templates.
It can also be seen that the effect of limb darkening is relatively modest, shifting the $v\sin{i}$ values systematically down by 1 km s$^{-1}$ if we lower the coefficient down to 0.5. This trend continues if we push the line limb darkening further down and in the extreme lowers $v\sin{i}$ by 3 km s$^{-1}$ (see also Shahbaz \& Watson 2007).  Since we expect the Roche-lobe filling donor star in its 9 day orbit around the neutron star to be closer to a giant rather than a main sequence dwarf, we use  $v\sin{i}$=32.9$\pm0.8$ km s$^{-1}$ in the remainder of this Letter. This is the mean value derived for our three giant templates while using a limb darkening of 0.75, as is appropriate for the R-band continuum in early K-giants. We remark that the bias introduced by this choice is very modest since all individual values are within $2\sigma$ of this and we will see below that the uncertainty on the derived neutron star mass is no longer dominated by the uncertainty in $v\sin{i}$, but instead by the radial velocity amplitude $K_2$. Very similar results were recently obtained by Shahbaz \& Watson (2007) using VLT echelle spectroscopy.

As mentioned above, these recent $v\sin{i}$ values are a factor of $\sim2$ lower than the previously reported value of Shahbaz et al. (1999).  Shahbaz \& Watson (2007) conclude that the dominant absorption blend near 6495\AA~ may have biased the 1999  $v\sin{i}$ determination which used data with relatively low signal to noise as well as low spectral resolution.
In our case, the high spectral resolution ensured that neither the instrumental profile of the spectrograph nor variable slit illumination effects had an impact on the analysis since the rotational broadening exceeded our resolution by a large amount. We also used templates that were obtained with the same instrument and had sufficient signal to noise to detect and resolve a large number of absorption lines. This permitted the determination of reliable values for $v\sin{i}$ with a better than 1 km s$^{-1}$ statistical precision. 

\subsection{The masses of the stellar components}

Armed with an improved determination of $v\sin{i}$, we can now derive new constraints on the mass of the compact object and its K-type donor. The measured radial velocity semi--amplitude $K_2$ and the rotational velocity give the mass ratio, $q=M_2/M_1$, of the system via $\frac{{\rm v\,\sin}i}{{\rm K}_2}=0.46[(1+q)^2q]^\frac{1}{3}$ \citep{1988ApJ...324..411W}. We can then solve for the masses of the stellar components using the mass functions which deliver $M\sin^3{i}$. Deriving actual masses requires knowledge of the binary inclination, $i$, which for V395 Car is thought to be close to 80 degrees (Mason et al. 1987). Values for $K_2$ have been reported in Jonker et al. (2005) which found $K_2$=99.1$\pm$3.1 km
s$^{-1}$ while Shahbaz et al. (2004) quote $K_2$=92.9$\pm$3.8 km
s$^{-1}$. 
This observed amplitude may be subject to a so-called K-correction due to the fact that the inner face of the companion star can be heated by X--ray emission coming from near the compact
object. The center of light measured via the spectral absorption features is
then offset from the true center of mass causing an overestimate of $K_2$. The magnitude of this effect depends on the irradiation geometry (e.g. \citealt{2005ApJ...635..502M}) as well as the intensity of the incoming radiation.
Shahbaz et al. (2004) argue that this correction is not significant in the case of their $K_2$, but Jonker et al. (2005) provide some evidence that X-ray heating does seem to be play a role in V395 Car. In the most extreme case of no absorption line contribution from the heated front face of the Roche lobe, the K-correction can be as large as $\Delta K<4/3\pi \times v \sin{i} = 14$ km s$^{-1}$.

\begin{figure}
\centerline{\includegraphics[angle=0,width=8.5cm]{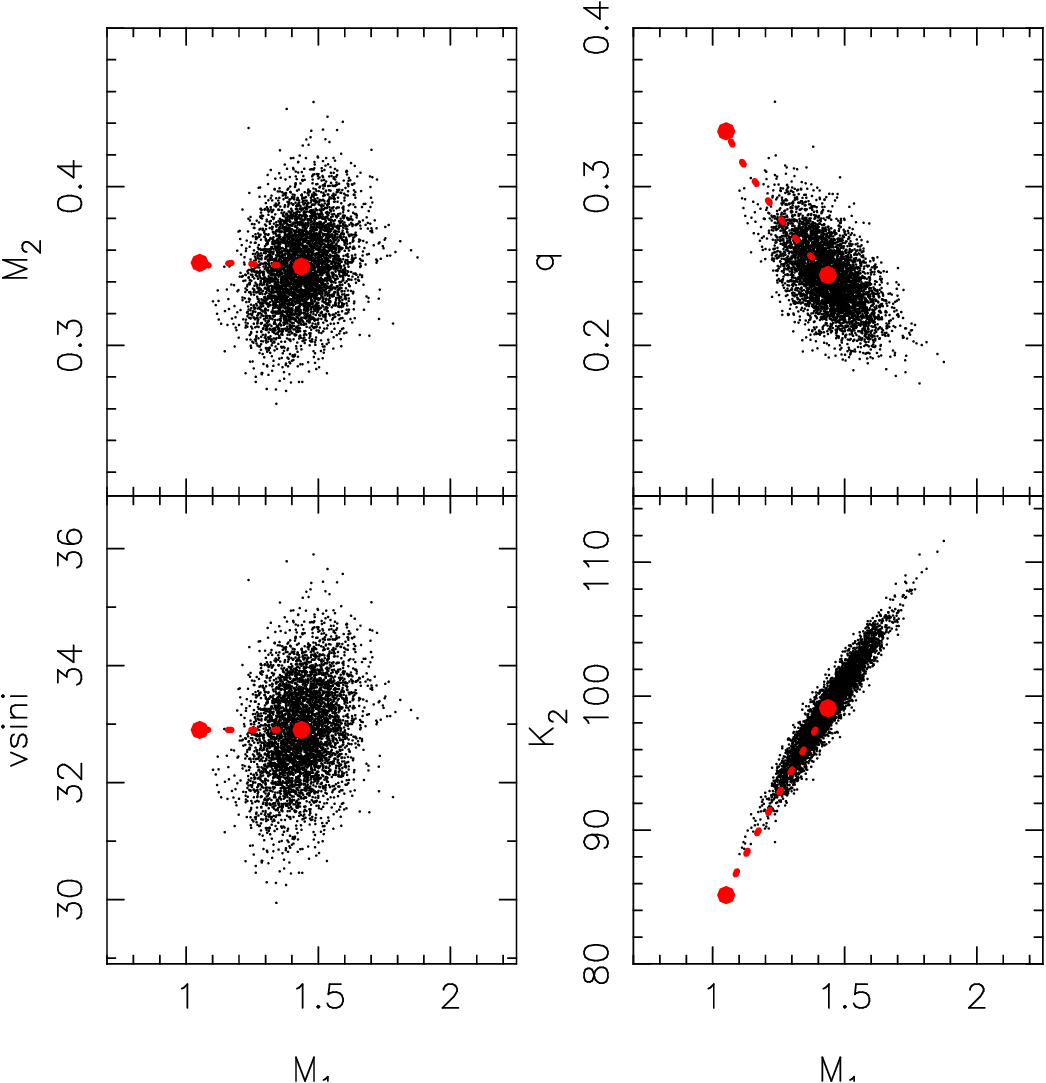}}

\caption{The four panels plot the distribution of system parameters for 5000 randomly selected combinations of $K_2$ and $v\sin{i}$. Clockwise starting at the upper left corner, we show the mass of the companion star,
the mass ratio $q$, the radial velocity amplitude
$K_2$ and $v\sin{i}$ plotted against the neutron star mass. The thick dashed curves show in which direction these probability clouds move when an irradiation correction to $K_2$ is applied with the solid circles indicating the system parameter solutions at the two extremes of no correction and maximum K-correction. \label{figmasses}}

\end{figure}

In order to determine the uncertainties on the component masses given the observables $K_2$, $v\sin{i}$ and $i$, we ran a Monte-Carlo simulation. The X-ray eclipse discussed in  Mason et al. (1987) gives us a good handle on the inclination. At face value, $K_2$=99.1 km s$^{-1}$ and $v\sin{i}$=32.9 km s$^{-1}$ gives a mass ratio of $q=0.25$. Using Table 2 in Mason et al. (1987) suggests that for this mass ratio, fits to the X-ray eclipse lightcurve imply an inclination angle of $\sim$83 degrees. Although somewhat model dependent, the inclination simply enters as $\sin^3{i}$, which is only a correction at the 2\% level for these high inclinations. Using $i=83^{\circ}$, we select random values for $K_2$ and $v\sin{i}$ by picking values from a normal distribution corresponding to the mean and $1\sigma$ error of the observed values. The system parameters are then calculated for each random set of parameters, and the process is repeated 5000 times. 
Figure \ref{figmasses} shows the corresponding probability clouds for various parameters. In order to illustrate the effect of the unknown $K_2$ correction due to X-ray heating, we plot the clouds for zero K-correction ($K_2$=99.1$\pm$3.1 km s$^{-1}$) and then show with dashed curves the steady trajectory the probability clouds would take as a function of the K-correction up to the maximally allowed correction.
We find that at the canonical values for $K_2$ and $v\sin{i}$, the neutron star mass is $M_1=1.44 \pm 0.10  M_{\odot}$, $M_2=0.35 \pm 0.03  M_{\odot}$ and $q=0.24 \pm 0.02$. The uncertainties reflect the $1\sigma$/68\% error values obtained by propagating the $1\sigma$ error values on the input parameters through our Monte-Carlo simulation. Our revised $v\sin{i}$ value has thus brought the neutron star mass down to a value remarkably close to the canonical neutron star mass. From the curves in Figure \ref{figmasses}, we see that the neutron star mass would go down further if any K-correction is applied with a minimum mass of $M_1=1.05 \pm 0.08  M_{\odot}$ at the maximal K-correction.
The mass for the donor star is also much lower than previously reported. The evolutionary timescale for such a low mass star is rather long and it is clear that single star evolution is not able to produce a $0.35M_{\odot}$ object that is able to fill its Roche-lobe in a 9 day orbit. This is not an uncommon finding in LMXBs and might be due to a phase of unstable and non-conservative mass transfer that caused a significant amount of mass to be lost from the donor.

We conclude that we have derived a revised estimate for the rotational broadening of the K-type mass donor star in the neutron star binary V395 Car. From this we derive new constraints on the mass of the neutron star which is most likely very close to the canonical neutron star mass of $1.4 M_{\odot}$. V395 Car is thus yet another case where a potentially massive neutron star appears to be brought back into line with the bulk of the known neutron stars.

\acknowledgments

We would like to thank Jeff McClintock and Eric Mamajec for acquiring the MIKE data at Magellan. The Magellan telescopes are operated at Las Campanas Observatory by the Magellan consortium consisting of the Carnegie Institution of Washington, Harvard University, MIT, the University of Michigan and the University of Arizona. The use of the spectral analysis software package \textsc {molly} written by Tom Marsh is acknowledged. DS acknowledges a STFC Advanced Fellowship as well as support through the NASA Guest Observer program.






\end{document}